\documentclass[a4paper,10pt]{article}
\usepackage[utf8]{inputenc}

\usepackage{amssymb}
\usepackage{graphicx}
\usepackage{epsfig}
\usepackage{psfrag}
\usepackage{latexsym}
\usepackage{indentfirst}
\usepackage{fancyhdr}
\usepackage{amssymb}
\usepackage{amsmath}
\usepackage{amsfonts}
\usepackage{colordvi}
\usepackage{pifont}
\usepackage{color}
\usepackage{graphicx}
\usepackage{url}
\usepackage{array}
\usepackage{rotating}
\usepackage{titling}

\title{Neutrino Mixing from $\Delta(6n^2)$ Groups}
\author{Thomas Neder $^{1\ast}$ \\{\small\it$^1$School of Physics and Astronomy, University of Southampton,}\\{\small\it Southampton, SO17 1BJ, U.K.}\\ {$^\ast$\footnotesize E-mail: \texttt{T.Neder@soton.ac.uk}}}
\begin{document}

\maketitle

\pagenumbering{gobble}

\begin{abstract}
Experimentally viable lepton mixing parameters can be predicted in so-called direct flavour models with Majorana neutrinos using $\Delta(6n^2)$ groups as a flavour group. In direct models, in which the flavour group is broken to a $Z_2\times Z_2$ subgroup in the neutrino sector, mixing angles and Dirac CP phase are purely predicted from symmetry. General predictions of direct models with $\Delta(6n^2)$ flavour groups are that all mixing angles are fixed up to a discrete choice and that the Dirac CP phase is $0$ or $\pi$; Furthermore, the middle column of the mixing matrix is trimaximal which yields the sum rule $\theta_{23}=45^\circ \mp \theta_{13}/\sqrt{2}$ depending on the Dirac phase. These predictions of lepton mixing parameters are compatible with recent global fit results or will be tested experimentally in the near future. It is the first time that such predictions have been obtained model-independently for an infinite series of groups.
\end{abstract}

\section*{Introduction}
The problem of the origin of neutrino masses and mixing is of fundamental importance and among the more specific questions that models of neutrinos have to attempt to answer are:
\begin{itemize}
 \item Why do particles mix as they do?
 \item Why are there 3 families of particles?
 \item Why are the masses of the particles what they are?
 \item Is there CP violation in the lepton sector?
\end{itemize}
A succesful and numerous class of models of neutrino masses and mixings are models with flavour symmetries: Generations of fermions are assigned to representations of an additional symmetry group called the flavour symmetry; for a review see e.g. \cite{King:2013eh}. With this additional symmetry, the full symmetry of the model becomes $G_\text{SM}\times G_\text{Flavour}$. The additional invariance of the Lagrangian under $G_\text{Flavour}$ restricts the allowed couplings in the Yukawa sector and eventually the allowed mass and mixing matrices. Often, the flavour symmetry needs to be broken, because the maximal flavour symmetry under which a Majorana mass term for three neutrinos with non-degenerate masses can be invariant is $Z_2\times Z_2$. This breaking can e.g. be achieved using scalar fields.

If the the flavour group is both broken to a $Z_2\times Z_2$ subgroup in the neutrino sector of the theory and to a discrete subgroup, often a cyclic group, in the sector of charged leptons, a model with a flavour symmetry is called \textit{direct}. In a direct model, the mixing angles and the Dirac CP phase are purely predicted from the choice of flavour group and its unbroken subgroups.

Many promising flavour symmetry candidates are $\Delta(6n^2)$ groups \cite{TooropJN,deAdelhartTooropRE,DingXX,KingIN,LamNG,KingAP,VarzielasSS,KrishnanSB,HolthausenVBA}. In particular, after the measurement of $\theta_{13}$ \cite{DayaBay,RENO,DCt13}, the only flavour symmetries in direct models that remain viable  are $\Delta(6n^2)$ groups or subgroups thereof \cite{d150lam,Holthausen:2012wt}. Furthermore, in direct flavour models, one can analyse $\Delta(6n^2)$ groups for all even n simultaneously \cite{KingVNA}. 

The groups $\Delta(6n^2)$ are non-abelian groups of order $6n^2$ and are isomorphic to a semidirect product: $\Delta(6n^2)\cong (Z_n\times Z_n)\rtimes S_3$ \cite{Escobar:2008vc}. The maybe best-known members of the series are $S_3=\Delta(6)$ and $S_4=\Delta(24)$. 

Any $Z_2\times Z_2$ subgroup of every $\Delta(6n^2)$ can give rise to a different mixing matrix. Consequently, all mixing matrices that are allowed by a certain $\Delta(6n^2)$ group can be found systematically by listing all $Z_2\times Z_2$ subgroups for each $\Delta(6n^2)$ group. In direct models with a $\Delta(6n^2)$ flavour group, only breaking $\Delta(6n^2)$ to $Z_3$ in the sector of charged leptons will produce experimentally viable predictions for mixing angles \cite{KingVNA}.

To produce concrete results for the mixing matrix, left-handed doublets are to transform (without loss of generality) under a 3-dimensional  representation with the generators (where $\eta=e^{2\pi i /n}$):
\begin{equation}
a=\begin{pmatrix}0&1&0\\0&0&1\\1&0&0\end{pmatrix},b=-\begin{pmatrix}0&0&1\\0&1&0\\1&0&0\end{pmatrix},c=\begin{pmatrix}\eta&0&0\\0&\eta^{-1}&0\\0&0&1\end{pmatrix},d=\begin{pmatrix}1&0&0\\0&\eta&0\\0&0&\eta^{-1}\end{pmatrix}.
\end{equation}
All $Z_2\times Z_2$ subgroups of a $\Delta(6n^2)$ group can be indexed by a single number which takes values $\gamma=1,\ldots,n/2$ \cite{KingVNA}. The columns of the mixing matrix are now given by the common eigenvectors of the elements of each $Z_2\times Z_2$ subgroup in the basis where the mass matrix of the charged leptons is already diagonal. Note that this approach does not determine the ordering of the columns and rows of the mixing matrix. The ordering of the mixing matrix can be fixed by demanding that the smallest entry be in the top right of the mixing matrix, i.e. $V_{13}$. With $\vartheta=\pi \gamma/n$ every mixing matrix has up to ordering the following form:

\begin{eqnarray}
V=\left(
\begin{array}{ccc}
 \sqrt{\frac{2}{3}} \cos (\vartheta ) & \frac{1}{\sqrt{3}} & \sqrt{\frac{2}{3}} \sin (\vartheta ) \\
 -\sqrt{\frac{2}{3}} \sin \left(\frac{\pi }{6}+\vartheta\right) & \frac{1}{\sqrt{3}} & \sqrt{\frac{2}{3}} \cos \left(\frac{\pi
   }{6}+\vartheta\right) \\
 \sqrt{\frac{2}{3}} \sin \left(\frac{\pi }{6}-\vartheta \right) & -\frac{1}{\sqrt{3}} & \sqrt{\frac{2}{3}} \cos \left(\frac{\pi }{6}-\vartheta
   \right) \\
\end{array}
\right).
\label{V}
\end{eqnarray}
This lepton mixing matrix is trimaximal with $\theta_{13}$ fixed up to a discrete choice and CP-phase 0 or $\pi$. For any value of $\vartheta$ the Dirac CP phase can be both $0$ or $\pi$ which corresponds to interchanging the two rows that are not fixed because they do not contain the smallest entry. This will also result in a different value for $\theta_{23}$. Generally, if the middle column is trimaximal, the following sum rule holds: $\theta_{23}=45^\circ\mp \theta_{13}/\sqrt{2}$ \cite{King:2013eh}.

\begin{figure}[h]
\center
\includegraphics[width=0.8\textwidth]{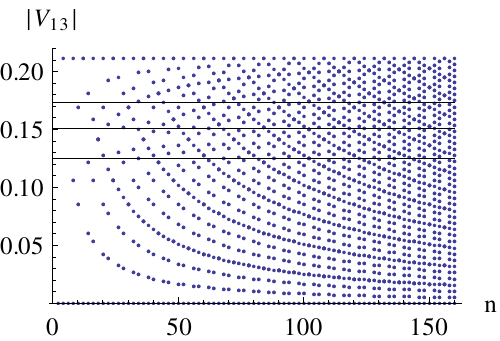}
\caption{Possible values of $|V_{13}|$; the lines denote the present approximate $3\sigma$ range of $|V_{13}|$ from \cite{CapozziCSA}. Examples include $|V_{13}|=0.211,0.170,0.160,0.154$ for $n=4,10,16,22$, respectively.  Each value of $V_{13}$ allows for two values of $\theta_{23}$ with $\delta_{CP}=0$ and $\delta_{CP}=\pi$ given by $\theta_{23}=45^\circ \mp \theta_{13}/\sqrt{2}$ respectively.}
\label{u13vsna}
\end{figure}

The allowed values of $|V_{13}|$ are plotted in figure (\ref{u13vsna}), where every choice of $\gamma$ and $n$ produces a dot. 
The number of allowed $Z_2\times Z_2$ subgroups and hence the number of allowed mixing matrices increases with $n$. 
If $n$ can be divided by 3, some of the subgroups produce the same mixing matrices which explains when dots seem to be missing. The horizontal lines denote the present approximate $3\sigma$ range of $|V_{13}|$ from \cite{CapozziCSA}. 

\section*{Conclusions and Outlook}
Direct models with a $\Delta(6n^2)$ flavour symmetry for Majorana neutrinos predict experimentally viable lepton mixing matrices. In \cite{KingVNA} lepton mixing predictions from $\Delta(6n^2)$ have been obtained for arbitrary even values of $n$ and all mixing matrices are shown to have the form of Eq. (\ref{V}). The predictions are compatible with current results of global fits and will be further tested in the near future. 

A direct model only predicts mixing angles and the Dirac CP phase from symmetry while Majorana phases remain unconstrained. Majorana phases can be predicted from symmetry in models with a generalised CP symmetry. For direct models with $\Delta(6n^2)$ flavour groups, generalised CP symmetries have been studied in \cite{KingRWA}. 
\section*{Acknowledgements}
TN acknowledges partial support from the European Union FP7 ITN-INVISIBLES (Marie Curie Actions, PITN- GA-2011- 289442).


\begin{thebibliography}{plain}
\bibitem{King:2013eh}
  S.~F.~King and C.~Luhn,
  Rept.\ Prog.\ Phys.\  {\bf 76} (2013) 056201
  [arXiv:1301.1340 [hep-ph]];
  S.~F.~King, A.~Merle, S.~Morisi, Y.~Shimizu and M.~Tanimoto,
  arXiv:1402.4271 [hep-ph].
  
  
\bibitem{TooropJN}
  R.~d.~A.~Toorop, F.~Feruglio and C.~Hagedorn,
Phys.\ Lett.\ B {\bf 703}, 447 (2011).
[arXiv:1107.3486 [hep-ph]].

\bibitem{deAdelhartTooropRE}
  R.~de Adelhart Toorop, F.~Feruglio and C.~Hagedorn,
Nucl.\ Phys.\ B {\bf 858}, 437 (2012).
[arXiv:1112.1340 [hep-ph]].

\bibitem{DingXX}
  G.~-J.~Ding,
Nucl.\ Phys.\ B {\bf 862}, 1 (2012).
[arXiv:1201.3279 [hep-ph]].

\bibitem{KingIN}
  S.~F.~King, C.~Luhn and A.~J.~Stuart,
Nucl.\ Phys.\ B {\bf 867}, 203 (2013).
[arXiv:1207.5741 [hep-ph]].

\bibitem{LamNG}
  C.~S.~Lam,
Phys.\ Rev.\ D {\bf 87}, no. 5, 053012 (2013).
[arXiv:1301.1736 [hep-ph]].

\bibitem{KingAP}
  S.~F.~King and C.~Luhn,
JHEP {\bf 0910}, 093 (2009).
[arXiv:0908.1897 [hep-ph]].

\bibitem{VarzielasSS}
  I.~de Medeiros Varzielas and G.~G.~Ross,
JHEP {\bf 1212}, 041 (2012).
[arXiv:1203.6636 [hep-ph]].

\bibitem{KrishnanSB}
  R.~Krishnan,
J.\ Phys.\ Conf.\ Ser.\  {\bf 447}, 012043 (2013).
[arXiv:1211.3364 [hep-ph]].

\bibitem{HolthausenVBA}
  M.~Holthausen and K.~S.~Lim,
Phys.\ Rev.\ D {\bf 88}, 033018 (2013).
[arXiv:1306.4356 [hep-ph]].


\bibitem{DayaBay}
  F.~P.~An {\it et al.}  [DAYA-BAY Collaboration],
  Phys.\ Rev.\ Lett.\  {\bf 108} (2012) 171803
  [arXiv:1203.1669];


\bibitem{RENO}
  J.~K.~Ahn {\it et al.}  [RENO Collaboration],
  Phys.\ Rev.\ Lett.\  {\bf 108} (2012) 191802
  [arXiv:1204.0626].



\bibitem{DCt13}
  Y.~Abe {\it et al.}  [Double Chooz Collaboration],
  arXiv:1301.2948 [hep-ex].

  
\bibitem{d150lam}
  C.~S.~Lam,
  Phys.\ Rev.\ D {\bf 87} (2013) 013001
  [arXiv:1208.5527 [hep-ph]].

\bibitem{Holthausen:2012wt}
  M.~Holthausen, K.~S.~Lim and M.~Lindner,
  Phys.\ Lett.\ B {\bf 721} (2013) 61
  [arXiv:1212.2411 [hep-ph]].
  

\bibitem{KingVNA}
  S.~F.~King, T.~Neder and A.~J.~Stuart,
Phys.\ Lett.\ B {\bf 726}, 312 (2013).
[arXiv:1305.3200 [hep-ph]].

\bibitem{Escobar:2008vc}
  J.~A.~Escobar and C.~Luhn,
  J.\ Math.\ Phys.\  {\bf 50} (2009) 013524
  [arXiv:0809.0639 [hep-th]].

\bibitem{CapozziCSA}
  F.~Capozzi, G.~L.~Fogli, E.~Lisi, A.~Marrone, D.~Montanino and A.~Palazzo,
[arXiv:1312.2878 [hep-ph]].


\bibitem{KingRWA}
  S.~F.~King and T.~Neder,
[arXiv:1403.1758 [hep-ph]].

\end{thebibliography}
\end{document}